# Superconductivity and weak anti-localization in nodal-line semimetal SnTaS$_2$

M. Singh[1], P. Saha[1], V. Nagpal[1] and S. Patnaik[1,*]

[1] School of Physical Sciences, Jawaharlal Nehru University, New Delhi, India

E-mail: spatnaik@mail.jnu.ac.in

## Abstract

Topological semimetals with superconducting properties provide an emergent platform to explore the properties of topological superconductors. We report magnetization, and magneto-transport measurements on high quality single crystals of transition metal dichalcogenide SnTaS$_2$. It is a nodal line semimetal with superconducting transition below T$_c$ = 2.9 K. Moderate anisotropy ($\gamma$ = 3.1) is observed in upper critical fields along $H$//c and $H$//ab plane. In the normal state we observe large magneto-resistance and weak anti-localization effect that provide unambiguous confirmation of topological features in SnTaS$_2$. Therefore, genuine topological characteristics can be studied in this material, particularly with regard to microscopic origin of order parameter symmetry.

# 1. Introduction

The role of electronic band topology in the condensed matter systems, particularly in topological superconductors, has attracted considerable attention in the recent past [1-3]. While the theoretical predictions on the characteristic signatures of such systems, are well documented [3,5,7,8], their experimental manifestations continue to be under persistent debate. In analogy with topological insulators, topological superconductors are predicted to be gapped superconductors in the bulk-interior along with Majorana fermions at the surface states [1-2,5,10,]. Realization of superconducting transition in such prototype systems via doping or external pressure has been reported [7,13,14]. Early topological superconductors were analyzed to possess topological surface states which was either due to proximity effects [8,15] or due to intercalation unto topological insulators [7,9,12,21]. However, unequivocal confirmation of the predicted p-wave singlet order parameter is still awaited [3,5]. Furthermore, detection of Majorana fermions has remained an open question. In this paper we report synthesis of high-quality single crystals of $SnTaS_2$ with superconducting $T_c$ of 2.9 K along with clear evidence for topological characteristics in the normal state.

The clearest signature of non-trivial topological band structure is the weak-anti localization (WAL) effect that can be studied using magneto-resistance measurements. This is primarily a quantum interference effect in the quantum diffusive regime [26,27]. In fact, this is frequently used as an alternative definition of topological insulators because of their delocalized surface states [26]. Over the years, several topological semimetals such as Dirac semimetal, Weyl semimetal, nodal-line semimetal have shown superconductivity with such non-trivial band structure. Some of them exhibit superconductivity in high pressure regime ($Cd_3As_2$, $WTe_2$) [13,14] while some others exhibit superconductivity at ambient pressures ($PbTaSe_2$, $PbTaS_2$). Moreover, superconductors having layered crystal structure are of interest due to their anisotropic properties and high critical fields that provide them similarity with the cuprates [11]. SnTaS$_2$ is one such layered transition-metal dichalcogenide where 2H-TaS$_2$ layers [16] are connected by intercalation of Sn atoms that makes a linear array of Sn-S-Sn units [17,18]. SnTaS$_2$ is reported to be a nodal line semi-metal [43] and superconducting below the critical temperature Tc~ 2.8K [19,20]. It is isoelectronic to PbTaS$_2$ [24] but it has centrosymmetric structure while PbTaSe$_2$ is non-centrosymmetric crystal. First principle calculations have shown that SnTaS$_2$ exhibits nodal-line band structure and drumhead like states when spin-orbit coupling (SOC) is excluded [24,25]. There have been some studies [23,24] on superconducting properties of SnTaS$_2$ but topological aspect related to non-trivial band structure is an open question. Here we address these issues by magneto-transport and magnetization measurements on high purity single crystals of SnTaS$_2$.

## 2. Experimental Technique

The single crystals of SnTaS$_2$ were synthesized using the chemical vapor transport method that consists of two parts. First, polycrystalline Sn$_{0.33}$TaS$_2$ was synthesized using the conventional solid-state reaction method by putting the stoichiometric mixture of Tin (Sigma Aldrich,99.8% pure), Tantalum (Alpha Aesar,99.9% pure) and Sulphur (Sigma Aldrich,99.98% pure) in an evacuated quartz tube. This was placed in a furnace at 850°C for two days. Then the polycrystalline sample was mixed with excess Sn powder to achieve elemental ratio of 1.2 :1: 2. It was vacuum sealed with iodine (3.5 mg/cc) in a quartz tube. Iodine was used as a transport agent. Excess Sn was mixed to rule out the reappearance of the Sn$_{0.33}$TaS$_2$ phase in the final product [24]. The tube was then put into a two-zone furnace (hot zone at 1000°C and cold zone at 960°C) for two weeks. Normal cooling to room temperature yielded shiny flake like single crystals of typical dimension ~ 2×2×0.02 mm$^3$. The structural characterization was done at room temperature using Rigaku X-ray diffractometer (XRD, Miniflex-600 with Cu-K$\alpha$ radiation). The single crystal XRD was done using Bruker x-ray diffractometer with Mo as X-ray source. The magneto-transport measurments were performed using a *Cryogenic* built

Cryogen Free Magnet (CFM) (8 Tesla, 1.6K) and a separate 14 Tesla *Cryogenic* Physical Property Measurement Systems (PPMS). Energy dispersive x-ray spectroscopy (EDAX) and Scanning electron microscopy (SEM) imaging were done using Bruker AXS microanalyzer along with a Zeiss EVO40 SEM analyser, respectively.

## 3. Results and Discussion

SnTaS$_2$ has a layered hexagonal structure that belongs to the space group (p6$_{3/mmc}$). The layered structure consists of alternating stacks of TaS$_2$ and Sn layers. [17,18]. Like 2H phase of TaS$_2$, SnTaS$_2$ is a centrosymmetric compound with the Sn layer held together between TaS$_2$ layers through Van der Waals bonding. XRD measurements on as grown crystals were carried out via two methods; on a flake-like thin crystal on a powder diffractometer as well as by collecting Laue data from a single crystal diffractometer. All the reflection peaks resemble with the underlying hexagonal symmetry that can be indexed in (0 0 *l*) direction [figure 1 (a)]. This is in agreement with c-axis oriented layered growth of SnTaS$_2$ single crystals [23]. Further, Laue pattern [shown in inset (ii) of figure 1 (a)] shows the concentric spot-like circles that reconfirms crystalline structure of single phase in the sample. The obtained XRD data were refined using Fullprof software to deduce the lattice parameters that come out to be a = b = 3.336 Å, c = 17.447 Å. This is in agreement with reported data [23,24]. The schematic unit cell of SnTaS$_2$ is shown in [inset (i) of figure 1 (a)]. A quantitative analysis of stoichiometric ratio was done using EDAX that was performed at several points on sample surface [figure 1(b)]. The atomic percent acquired through EDAX is close to the intended stoichiometric values; that Sn : Ta : S comes out to be 30%, 23%, 47% (Sn : Ta : S). The scanning electron microscope image [inset (ii) of figure 1 (b)] clearly shows the layered morphology of the specimen.

The measurements of temperature dependent electrical resistivity were carried out using linear four probe method. Figure 2 (a) shows the zero field resistivity data from 1.6 K to

300 K. Inset (i) of figure 2 (a) shows the evidence for superconducting transition at ~ 2.9 K. The transition is quite sharp and its transition width is ($\Delta T_c = Tc_{onset} - Tc_{zero} = 0.24$ K). The RRR (Residual Resistivity Ratio = $\rho(300K)/\rho(4K)$) is estimated to be 530, which is the highest RRR reported in this system [23,24]. It reflects high quality of the single crystal that is devoid of defects and impurities. Furthermore, a strikingly linear behavior, with intercept passing through the origin, is seen in the normal state temperature dependence of electrical resistivity. This is akin to optimally doped cuprates [11,53]. In general terms, electrical resistivity of metals exhibits such linear behavior when the excitonic (phononic or spin waves etc.) energy scale is lower than $k_BT$. At lower temperatures, still in the normal state, there would be a crossover to a region where excitonic scale would be higher than thermal energy. This is reflected as a power law behavior of resistivity. The normal state resistivity data therefore could be divided into 3 parts. First part is from just above the $T_c$ to around 40K, where resistivity curve shows power law behavior. Second region is from 40K to 140K with a slope of 1.83μΩ-cm K$^{-1}$ is seen. The third region shows linear behavior with a slope 1.55 μΩ-cm K$^{-1}$ from 140K upto 300K with intercept passing through the origin.

Next we discuss the dc magnetization measurement that was performed in *H //* ab and *H //* c directions of the crystal. Zero Field Cooled (ZFC) and Field Cooled (FC) data were taken along *H//*ab direction using field of 1mT which is shown in [inset (ii) of figure 2(a)]. The critical temperature obtained from magnetization measurements (2.9±0.1K) matches well with the $T_c$ obtained from resistivity measurements. Superconducting shielding volume fraction of the sample deduced from ZFC susceptibility comes out to be (99%). The magnetization versus magnetic field (M-H) loop for both orientation (*H//*ab) and (*H//*c) is shown in figure 2 (b) and inset of figure 2 (b) respectively. For the determination of lower critical filed $H_{c1}$ in *H//*ab direction, the M-H data at different temperature were taken in the fourth quadrant. This is shown in figure 2(c). Deviation from the linear diamagnetic behavior extrapolates to the lower critical field at each temperature. Figure 2(d) shows fitting between $H_{c1}$ as a function of temperature using the parabolic relation $H_{c1,ab}(T) = H_{c1,ab}(0) (1 – t^2)$, where t denotes reduced temperature $T/T_c$. Since the demagnetization effects are negligible in H//ab direction (due to thin platelike sample) lower critical field can be calculated with sufficient accuracy. The extrapolation of the curve leads to determination of $H_{c1, ab}(0)$ as 4.7±0.2 mT. From the M-H along the H//c direction, taken at 1.6 K [figure 2(b)], the obtained value for $H_{c1,c}$ comes to be 0.58mT. Using similar extrapolations, $H_{c1,c}(0)$ is estimated to be 1.88mT. Since demagnetization factor in *H*//c orientation loop can't be neglected, we calculated the actual value after demagnetization factor correction by using the Brandt's formula [30], $H'_{c1,c}(0) = H_{c1,c}(0)/(tanh(0.67c/a)^{1/2})$ where c and a are the thickness of the sample and length perpendicular to the field direction respectively. The demagnetization corrected $H_{c1,c}$ comes out be 20.5 mT.

Next, we discuss the onset of magneto-resistance in the presence of different magnetic fields both in the orientation of (*H*//c) and (*H*//ab) directions [figure 3(a) and 3 (b)]. The superconducting transition gets broadened when applied field is increased, more so when the field is applied parallel to c-axis of the crystal. Critical temperature for various fields is obtained from the mid-point of in-field transition criterion in both the orientations. In *H*//c direction the $H_{c2}$ versus temperature curve [inset of figure 3(a)] can be fitted using the Ginzburg-Landau equation $H_{c2,c}(T) = H_{c2,c}(0) (1-t^2)/(1+t^2)$, where $t = T/T_c$. The data fit with the GL equation very well and the intercept of the fit at y-axis gives the value of upper critical field $H_{c2,c}(0)$. This is estimated to be 25.2±0.9mT. The Pauli limit of upper critical field in weak coupling limit is given by $1.84T_c = 5.3T$, which indicates orbital effects limit the upper critical field magnitude in $SnTaS_2$. The temperature dependence of upper critical field in *H*//ab direction, shows upward behavior which results in deviation from the GL fit [inset of figure 3(b)]. In such cases $H_{c2,ab}$ can be fitted using the equation $H_{c2,ab}(T) = H_{c2,ab}(0) (1-t^{5/2})^{5/2}$ [23].. The upward curvature in upper critical field (near $T_c$) seen here has been previously reported in SnTaS$_2$ [23,24], PbTaSe$_2$ [47,48] and other intercalated chalcogenides [39,41]. There are several possible origin for this upward curvature in upper critical field in $H//ab$ direction. The primary cause could be the effect of impurities and disorders [34,35], but this could be discarded on the basis of high RRR of single crystals reported in this study. Other reasons are dimensional crossover [31], multiband effect and non-local effects in clean limit [32]. Previous reports in organic molecules [36,38] and tantalum based intercalated dichalcogenides [39,41] have explained this upward behavior with dimensionality crossover model [31]. Along with that Pauli paramagnetic limit is also exceeded in such cases. But as the coherence length $\xi_c$ in the case of SnTaS$_2$ is larger than the interlayer spacing (~8.72Å), the bulk behavior is indicated. Further, we note that H$_{c2,ab}$ (79.1±0.016 mT) is less than the Pauli limit. Hence the dimensional crossover model is not suitable to explain upward curvature in upper critical field. By deduction, this implies that possible non-local effects may be the dominant reason for the departure from GL behavior. Furthermore, the anisotropy parameter calculated as the ratio of two upper critical fields i.e. [$\gamma$ = H$_{c2,ab}$(0)/H$_{c2,c}$(0)] yields the value of 3.1. This is close to the value obtained from H$_{c1}$ measurements and is substantially lower than previous reports [23,24]. The Ginzburg-Landau (GL) coherence lengths in both directions $\xi_{ab}$ and $\xi_c$ are calculated using the formulas $\xi_{ab} = (\Phi_0/2\pi H_{c2,c})^{1/2}$ and $\xi_c = \Phi_0/2\pi \xi_{ab} H_{c2,ab}$ respectively [28,29] with $\Phi_0$ being the magnetic flux quantum. These calculations lead to estimation of coherence length as $\xi_{ab}$ = 114.3 nm and $\xi_c$ = 36.4 nm. The GL parameters $\kappa_{ab}(0)$ and $\kappa_c(0)$ can be calculated by using the formula $\kappa_{ab}(0) = (\lambda_{ab} \lambda_c / \xi_{ab} \xi_c)^{1/2}$ and $\kappa_c(0) = \lambda_{ab}/\xi_{ab}$. From the ratio of upper critical field and lower critical field, $\kappa_{ab}$ can be obtained, because H$_{c2,ab}$/H$_{c1,ab}$ = $2\kappa^2_{ab}$/(ln $\kappa_{ab}$) [21,24]. Therefore $\kappa_{ab}$ = 3.039. This further leads to determination of penetration depth $\lambda_{ab}(0)$ = 110.6 nm and $\lambda_c(0)$ as 347.3 nm.

Next, we turn to the question as to whether the superconducting state in $SnTaS_2$ is derived from non-trivial topological states as indicated in the reported electronic band structure [25]. $SnTaS_2$ is a nodal line semimetal in the normal state and it is worthwhile to study its topological properties. Such materials have the unique surface states that make them different from trivial class of materials [1-3]. In topological insulators the surface states are helical Dirac states and the upspin and downspin components are delocalized on opposite surfaces intertwined by the bulk states. From transport measurements this is ascertained from Weak anti-localization theories [26,27,42,45]. The experimental manifestation of this is the negative magnetoconductivity that is the decrease in electrical conductivity in the presence of external magnetic field. This is because in the quantum diffusion regime, the time reversed inverted scattering trajectories interfere destructively, giving rise to enhanced conductivity [26]. External magnetic field can destroy these interference effect which can give rise to negative magneto-conductivity. Although nodal-line semimetals in general do not possess surface states, the trajectory along the nodal line under adiabatic limit is such that it picks a $\pi$ Berry phase [43]. By definition, there is a minimal overlap of conduction and valence band in a semimetal where the 1D loop constitute the nodal line [Inset (i), Figure 4]. If the Fermi level is not coincident with the crossover loop then the Fermi surface becomes toroidal. In single crystals, the short range disorder driven scattering will be relatively negligible compared to long range scattering mechanism such as Coulombic scattering etc. Since the scattering momentum along the circumferential direction would be negligible, the cross-sectional loop would be dominant with $\pi$ Berry phase. Recent theoretical reports have claimed the Weak localization (WL) and weak anti-localization (WAL) feature in nodal-line semimetals depending on the length scale of potential responsible for scattering [44,51]. Inset (ii) of Figure 4 shows magneto transport measurement up to ±5T at 4K in the normal state of single crystal of $SnTaS_2$. A large non-saturating magnetoresistance of 320% is estimated by using the formula [(R(B)-

R(0))/R(0)]×100%. This non saturating MR can be attributed to topological feature because classical MR saturates at high fields [41,46]. A special feature observed is the V-shape cusp like behavior at low fields up to ±1.2T [figure (4)] [49,50]. This cusp like behavior confirms the WAL effect in SnTaS$_2$. There is the predicted increase in conductivity at lower temperatures [ inset (iii) figure (4)] and negative magneto-conductivity [26,27,42] [Figure (4)]. The large WAL effect in SnTaS$_2$ can be understood qualitatively as follows. In weakly disordered nodal-line semimetals, the screening effect becomes unconventional that means the scattering potentials becomes long ranged [51]. In such limit, motion of quasiparticles can be confined to 2D planes perpendicular to the nodal line, and backscattering is dominated by those loops which encircle the nodal line [44].

In essence the additional π Berry phase of nodal line makes the interference destructive leading to WAL. This effective 2D diffusion is therefore the reason which makes the WAL correction so large since there are a large number of such 2D subsystems [44]. Conductivity of each subsystem adds up to give such a large conductivity. Further if the Fermi energy above the nodal line is low, WAL effect become favorable [44]. Previous first principle calculations [24,25] as well as ARPES studies [25] have reported that the nodal lines in SnTaS$_2$ is in the vicinity of Fermi level. So SnTaS$_2$ satisfies all the requirements for large WAL effect confirming dominance of topological features in the normal state.

## 4. Conclusion

In summary we report the synthesis of high-quality single crystal of nodal line semimetal SnTaS$_2$ which is intercalated transition metal dichalcogenide. The specimen was grown using chemical vapor transport method. The residual resistive ratio comes out to be ~530 with no signatures of structural phase transition. The onset resistive critical temperature is found to be 2.9±0.1K. The upper critical field H$_{c2,c}$ fits very well with GL fit while H$_{c2,ab}$ shows upward behavior neat $T_c$. Anisotropy factor is estimated to be 3.1. SnTaS$_2$ is a Pauli limited weakly coupled Type-II superconductor. Magneto-transport measurements shows a non-saturating MR of 320% at 4K up to ±5 Tesla. We observe WAL effect at low fields which is first ever experimental confirmation of WAL in SnTaS$_2$. In essence, SnTaS$_2$ provides a good platform for understanding the superconducting transition metal dichalcogenides with nodal-line fermions. A lot about the pairing symmetry in the superconducting state as well as the topological feature need to be unraveled in this rekindled superconducting system.

**Acknowledgements**

M. Singh acknowledges CSIR for support through CSIR-JRF fellowship. P. Saha, V. Nagpal thank UGC for providing JRF. We thank FIST program of Department of Science and Technology, Government of India for low temperature high magnetic field facility at JNU. We thank Advanced Instrumentation Research Facility (AIRF), JNU for technical support.

**Figure Captions:**

**Figure 1** (a) X-ray diffraction data of single crystal $SnTaS_2$. inset (i) of (a) shows the schematic unit cell of $SnTaS_2$. Inset (ii) shows Laue's pattern. (b) EDX mapping of $SnTaS_2$ crystal. Inset (i) shows the optical micrograph of single crystal and inset (ii) of figure 1(b) shows SEM image of crystal which confirms the layered morphology.

**Figure 2** (a) Temperature dependent resistivity from 1.6K up to 300K. inset (i) shows the superconducting transition. Inset (ii) is the dc magnetization (ZFC and FC) along $H//$ab direction. (b) MH loop is plotted for H//c direction. Inset (i) shows the MH loop along for $H//ab$ plane. (c) Zero field cooled magnetization curves at different temperatures along *ab* plane (T = 1.75K, 1.9K, 2.2K, 2.4K, 2.6K). (d) Estimated lower critical field points and parabolic fit for $H_{c1}(T)$ along *ab* plane.

**Figure 3** The low temperature resistivity is plotted at at different fields (a) in H//c direction, at (H = 0mT, 2mT, 4mT, 6mT, 8mT, 10mT) and (b) along H//ab plane at (H = 0mT, 5mT, 10mT, 15mT, 20mT, 30mT). Inset (i) of figure 3(a) shows the extrapolated $H_{c2}(T)$ fit with Temperature (Ginzburg-Landau upper critical field) in H//c direction. Inset (i) of figure 3(b) shows the upper critical field $H_{c2}(T)$ fit for H//ab plane.

**Figure 4** The negative magnetoconductivity due to WAL effect in $SnTaS_2$. Inset (i) shows the schematic band diagram of a nodal line semimetal and a torus shaped Fermi surface with the path (Blue color) which encloses π Berry phase. Inset (ii) shows the transverse MR up to ±5 Tesla. Inset (iii) shows the conductivity (σ) curve at low temperatures (from 4.5K to 10K).

**Figure 1**

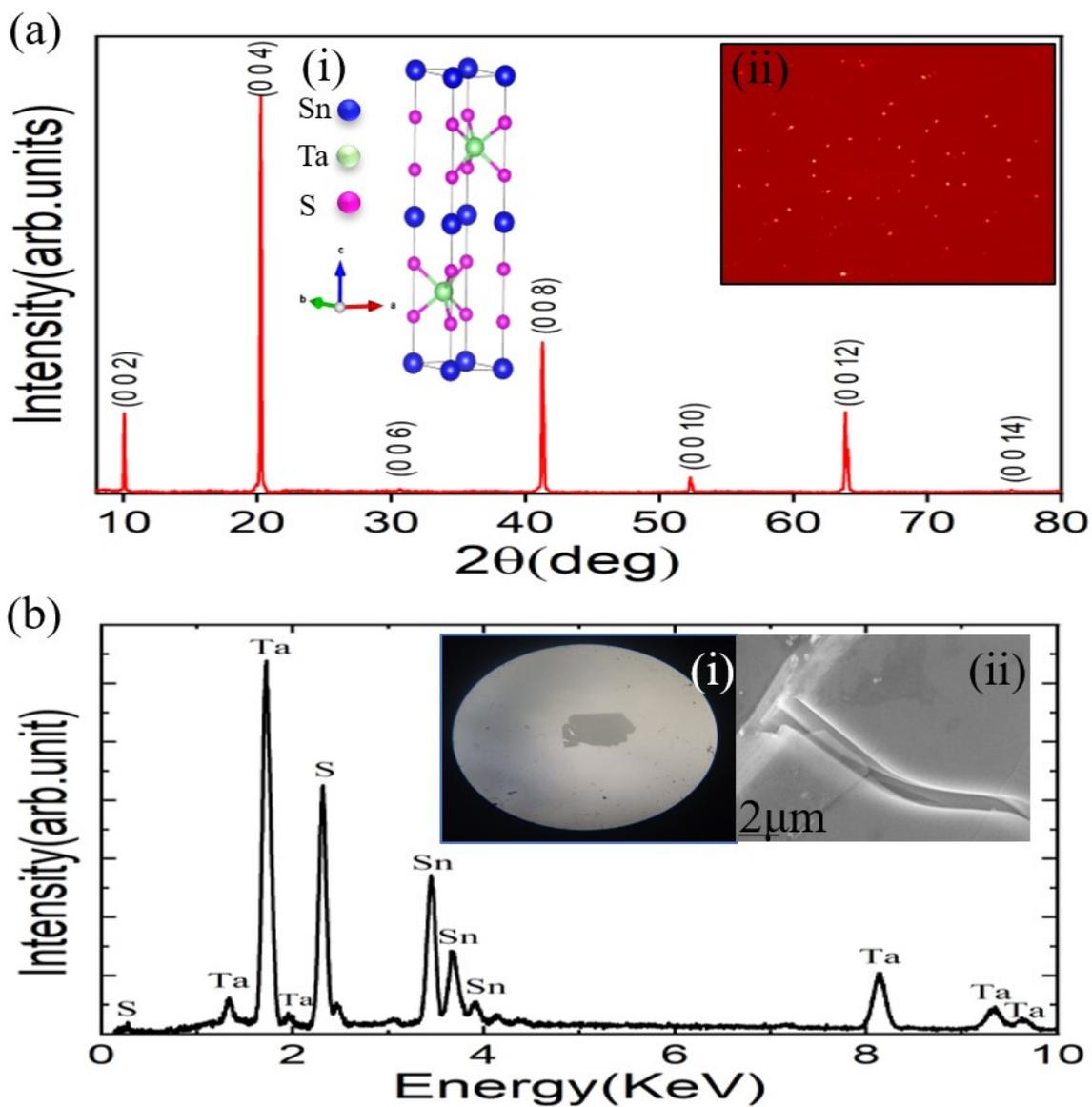

**Figure 2**

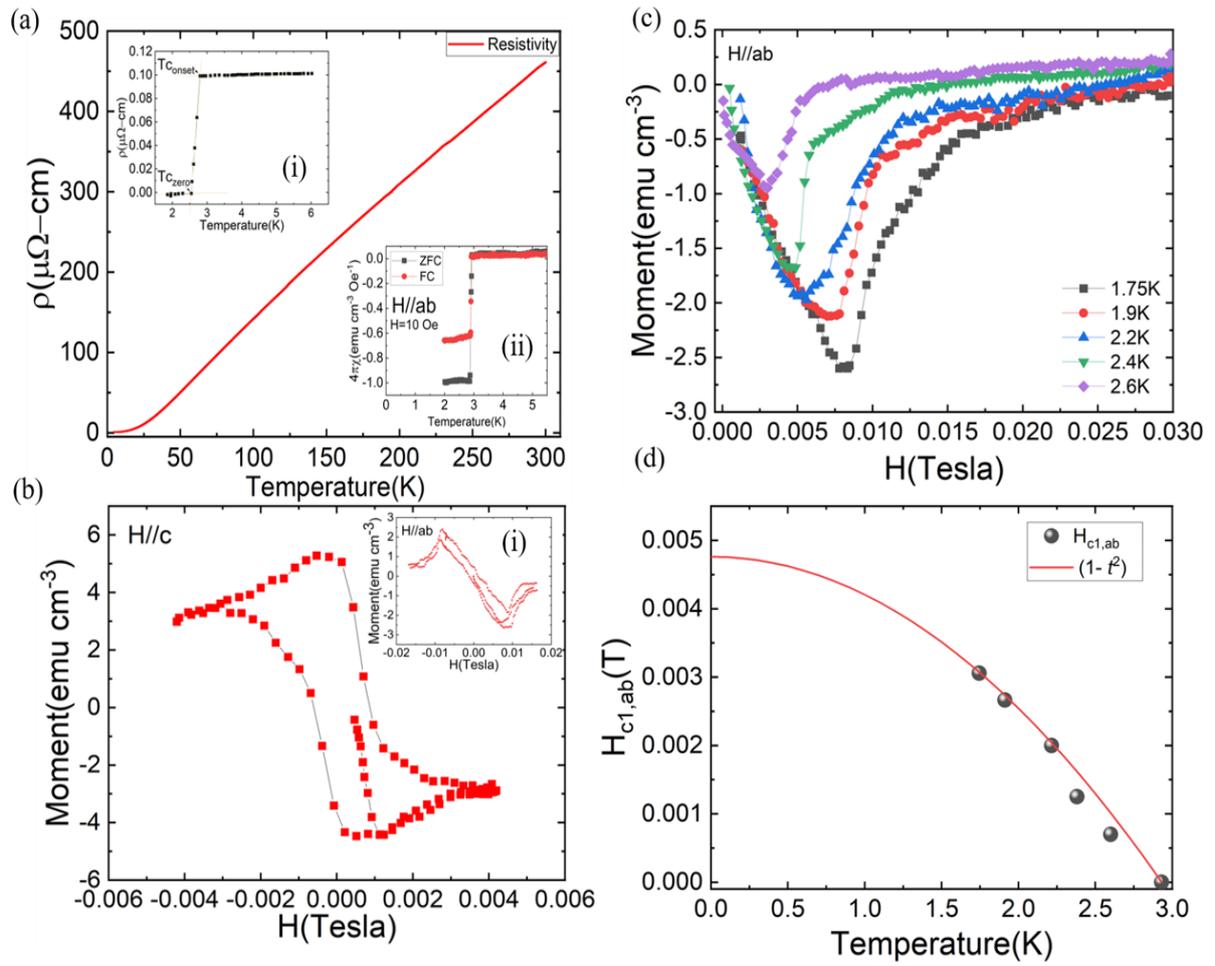

**Figure 3**

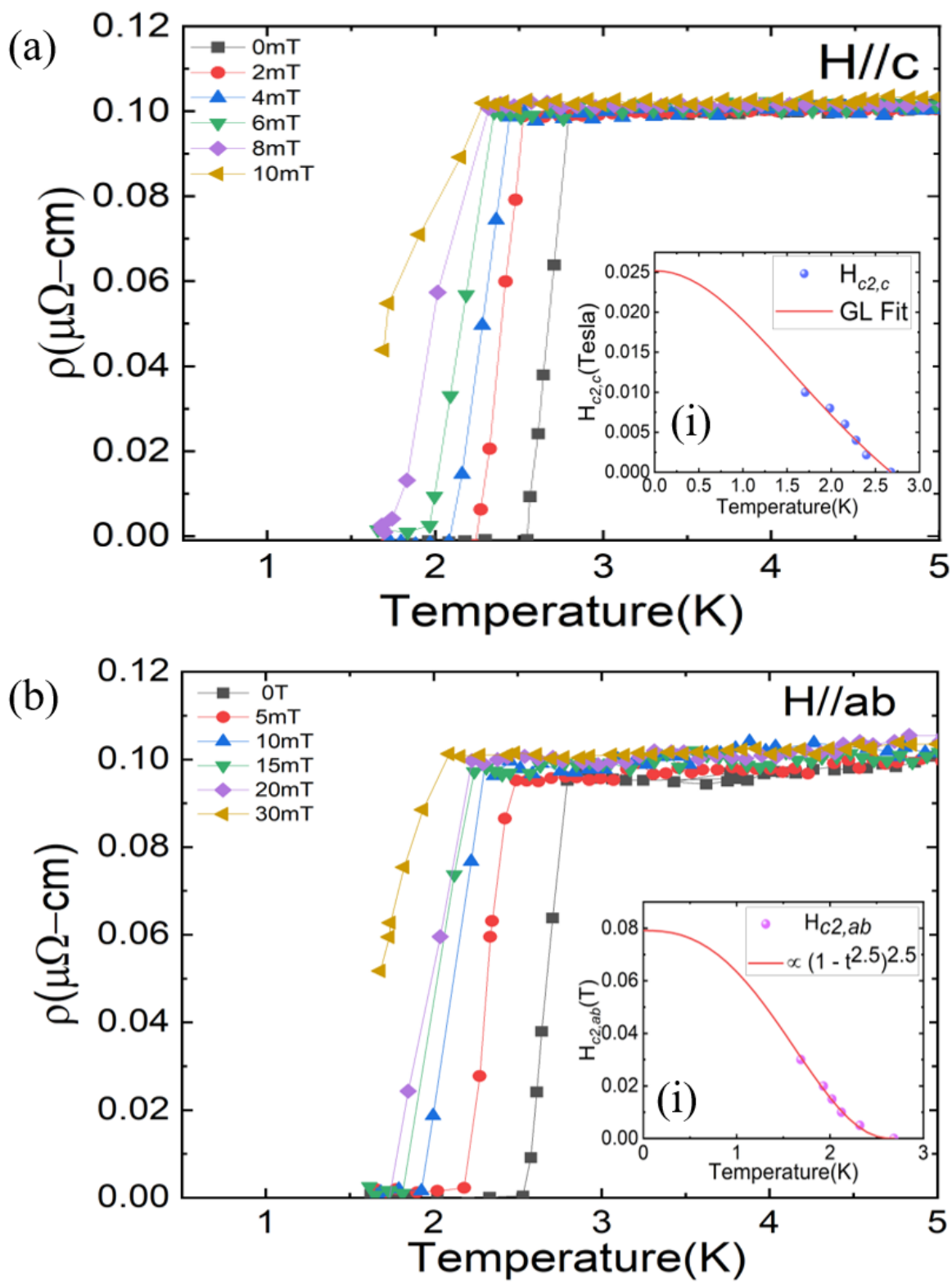

**Figure 4**

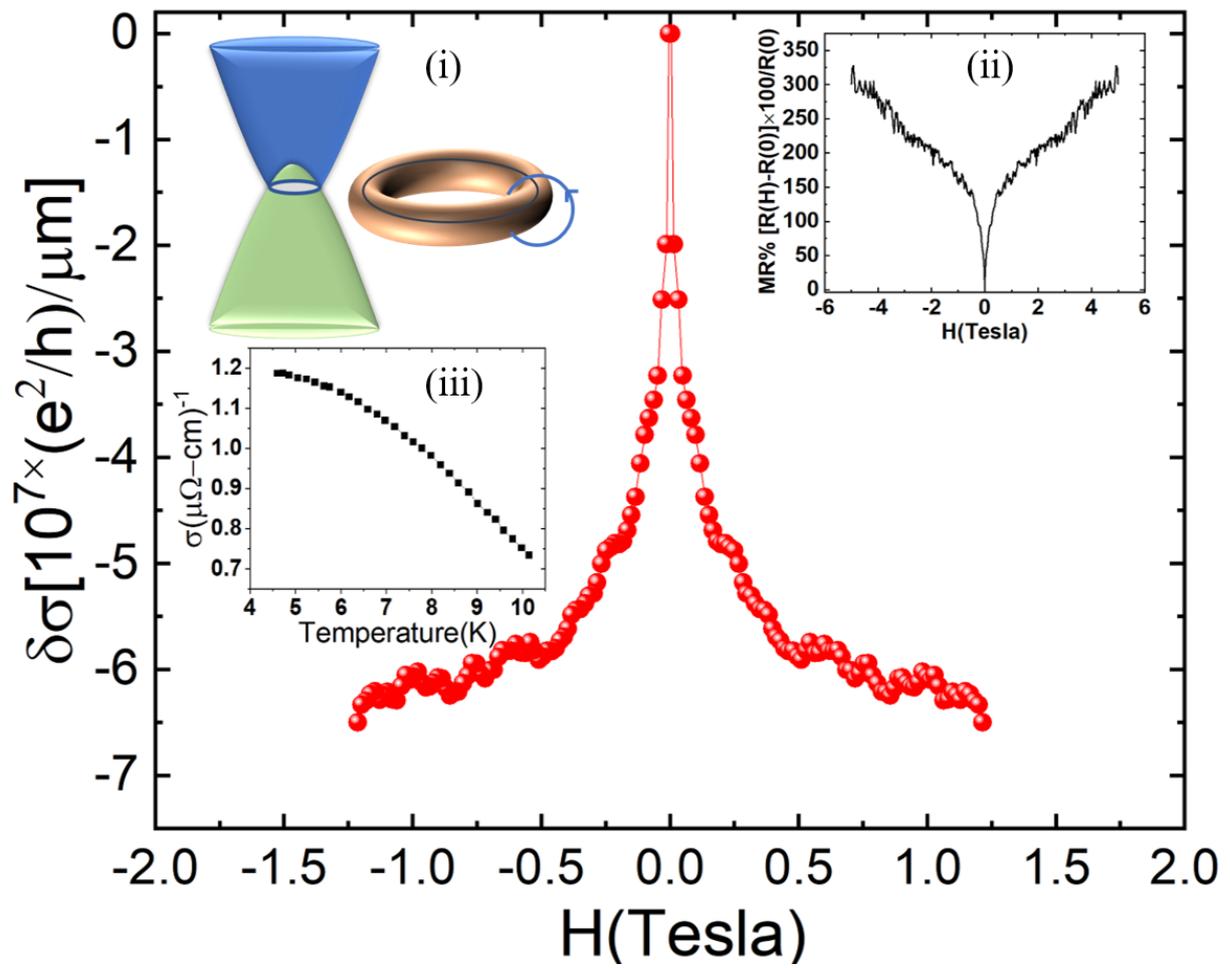